\documentclass[12pt,makeidx,epsfig,floatfig,wrapfig,amsbsy,amsfonts]{article}
\usepackage{times}
\usepackage{amsmath}
\usepackage{graphicx}
\usepackage{epsfig}
\usepackage{wrapfig}
\advance \topmargin by -\headheight
\advance \topmargin by -\headsep
\evensidemargin \oddsidemargin
\begin{document}
\pagestyle{plain}
\begin{titlepage}
\flushright{IHEP 2005-20}
\flushright{\today}
\flushright{OEF}
\vspace*{0.15cm}
\begin{center}
{\Large\bf
      Observation of the radiative kaon decay \\
  K$^{-} \rightarrow \mu^-\pi^{\circ}\gamma\nu   $  \\
    }
    
\vspace*{0.15cm}

\vspace*{0.3cm}
{\bf  
O.G.~Tchikilev, S.A.~Akimenko,  
 G.I.~Britvich, K.V.~Datsko,  A.P.~Filin, 
A.V.~Inyakin,  V.A.~Khmelnikov, A.S.~Konstantinov, I.Y.~Korolkov,  V.M.~Leontiev, V.P.~Novikov,
V.F.~Obraztsov,  V.A.~Polyakov, V.I.~Romanovsky, 
  V.I.~Shelikhov,        V.A.~Uvarov,   O.P.~Yushchenko. }
  
\vskip 0.15cm
{\large\bf $Institute~for~High~Energy~Physics,~Protvino,~Russia$}

\vskip 0.35cm
{\bf 
 V.N.~Bolotov, S.V.~Laptev,   V.A.~Duk, A.Yu.~Polyarush.}
\vskip 0.15cm
{\large\bf $Institute~for~Nuclear~Research~Moscow,~Russia$}
\vskip 0.15cm
\end{center}
\end{titlepage}
\begin{center}
Abstract
\end{center}

 Using data collected with the ``ISTRA+'' spectrometer during the 2001 run
 of the U-70 proton synchrotron in Protvino, we report the first observation
 of the radiative kaon decay K$^{-} \rightarrow \mu^{-}\pi^{\circ}\gamma\nu $.  
 We find BR(K$_{\mu3\gamma}, 5<$E$^*_{\gamma} < 30~$ MeV) / BR(K$_{\mu3}$) =
 0.270 $\pm$ 0.029(stat) $\pm$ 0.026(syst)\% and
  BR(K$_{\mu3\gamma}, 30<$E$^*_{\gamma} < 60~$MeV$)/$BR(K$_{\mu3}) =
  0.0448 \pm 0.0068($stat$) \pm 0.0099($syst$)\%$.
   These ratios are consistent with
  the  theoretical predictions $0.21\%$ and $0.047\%$ respectively. The measured
  angular distribution asymmetry for the region $5<$E$^*_{\gamma} <30$~MeV, 
   A$(\cos \theta^*_{\mu\gamma})= 0.093 \pm 0.141$,
  is two standard deviations away from the 
   theoretical prediction of 0.354. The measured asymmetry in
  the T-odd variable   $\xi = \vec p_{\gamma} \cdot ( \vec p_{\mu} \times \vec p_{\pi})/m^3_K$
  is   $-0.03 \pm 0.13$.

\vskip 0.35cm

\large

\section{ Introduction}

  The study of the radiative kaon decays can give valuable information on the kaon
  structure and allows for good test of theories describing hadron interactions and
  decays, like Chiral Perturbation Theory (ChPT). Until now the studies of the
  radiative K$_{l3}$-decays are restricted by the decay modes with electrons in the
  final state or by the studies of K$_L$ decays~\cite{na48,ktev2}. Only one paper~\cite{limit}, 
  dated by 1973,
   published the upper limit on the branching of K$^+_{\mu3\gamma}$-decay.
  
  The interest to the study of K$_{l3\gamma}$ decays is further enhanced by the theoretical
  proposals to search for effects of new physics using T-odd kinematical variable
  $\xi=\vec p_{\gamma}\cdot(\vec  p_{l}\times\vec p_{\pi})/m^3_K$~\cite{braga1,braga2}.
  In the standard model the expected asymmetry for K$^-_{\mu3\gamma}$ decay
\begin{center}  
   A$(\xi) = \frac{N(\xi >0) -N(\xi <0)}{N(\xi >0) + N(\xi <0)}$
\end{center}   
  is at the level $1.14 \times 10^{-4}$~\cite{braga1}, whereas in the extensions of the
  standard model it can achieve $2.6 \times 10^{-4}$~\cite{braga2}.
   
  In this paper we present first observation of the  
  radiative K$^-_{\mu3}$~~ decay.  The experimental setup and event selection are described
 in section~2. The results of the analysis are presented in section~3, where 
 we,  first, show the presence of the signal for the photon energy in the kaon rest frame
 E$^*_{\gamma}$ below 60~MeV,   measure  the branching
  ratio for the region 5 $<$ E$^*_{\gamma}$ $<$ 30~MeV,  measure the asymmetries
  in this region and finally 
  measure  the branching ratio for the region
   30 $<$ E$^*_{\gamma}$ $<$ 60~MeV.
  Our conclusions are given in the last section.

\section{ Experimental setup and event selection}

\begin{figure}
\epsfig{file=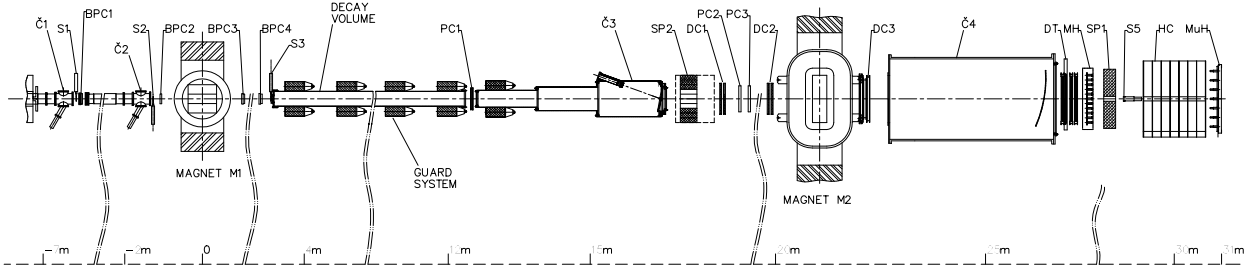,height=16.5cm,angle=90}
\caption{The elevation view of the ISTRA+ setup. M$_1$ and M$_2$ are magnets,
\v{C}$_i$ --- Cerenkov counters, BPC$_i$ --- beam proportional chambers,
 PC$_i$ --- spectrometer proportional chambers, SP$_i$ --- lead glass
 calorimeters, DC$_i$ --- drift chambers, DT$_i$ --- drift tubes,
 HC --- a hadron calorimeter, S$_i$ --- trigger scintillation counters,
 MH --- a scintillating hodoscope and MuH --- a scintillating
 muon hodoscope.}
\end{figure}

The experiment is performed at the IHEP 70 GeV proton synchrotron U-70.
The    ISTRA+ spectrometer has been described in detail in 
recent papers on $K_{e3}$  \cite{ref2,ref2n}, $K_{\mu 3}$ \cite{ref3,ref3n} 
  and $\pi^-\pi^{\circ}\pi^{\circ}$ decays \cite{ref4}.
Here we recall briefly  the characteristics relevant to our analysis. 
 The  ISTRA+ setup is located in a negative unseparated 
 secondary beam line 4A of U-70. The beam momentum  is $\sim 25$ GeV/c with 
$\Delta p/p \sim 1.5 \%$. The admixture of $K^{-}$ in the beam is $\sim 3 \%$,
 the beam intensity is $\sim 3~\cdot~10^{6}$ per 1.9 sec U-70 spill.
 A schematic view of the ISTRA+ setup is shown
 in Fig.~1.
  The  beam particles are deflected by the magnet M$_1$
 and are measured by four proportional chambers BPC$_1$---BPC$_4$ with 1~mm
 wire spacing, the kaon identification is done by three threshold Cerenkov
 counters \v{C}$_0$---\v{C}$_2$. The 9~meter long vacuum decay
 volume is surrounded by eight lead glass rings used to veto low energy
 photons. The  72-cell lead-glass calorimeter SP$_2$ plays the same role.
 The decay products are deflected in the magnet M$_2$ with 1~Tm field integral
  and are measured with 2~mm step proportional chambers PC$_1$---PC$_3$, with
 1~cm cell drift chambers DC$_1$---DC$_3$ and, finally, with 2~cm diameter
 drift tubes DT$_1$---DT$_4$. The wide aperture threshold Cerenkov counters
 \v{C}$_3$~,\v{C}$_4$~, filled with He,  serve to trigger electrons
 and are not used in the present measurement. SP$_1$ is a
 576-cell lead-glass calorimeter, followed by HC, a scintillator-iron
 sampling hadron calorimeter. MH is a 11x11 cell scintillating hodoscope,
 used to  improve the time
 resolution of the tracking system, MuH is a 7x7 cell muon hodoscope.
 
  The trigger is provided  by scintillation counters S$_1$---S$_5$,
  beam Cerenkov counters and by the analog sum of amplitudes from
  last dynodes of the SP$_1$ : 
   T=S$_1 \cdot \mbox{S}_2 \cdot \mbox{S}_3 \cdot \overline{\mbox{S}}_4\cdot 
   \mbox{\v{C}}_1
   \cdot \overline{\mbox{\v{C}}}_2 \cdot 
   \overline{\mbox{\v{C}}}_3 \cdot
   \overline{\mbox{S}}_5 \cdot \Sigma(\mbox{SP}_1)$,
   here S$_4$ is a scintillation counter with a hole
   to suppress the beam halo, S$_5$ is a counter downstream of the setup at the
   beam focus, $\Sigma$(SP$_1)$   requires that the analog sum to be
   larger than the MIP signal.

 During the  physics run in November-December 2001 350 million trigger
 events were collected with  high beam intensity. This information
 is complemented by 150~M  Monte Carlo~(MC) events generated using 
 Geant3 \cite{ref5}
 for the dominant $K^-$ decay modes, 100~M of them are the mixture of the dominant
 decay modes with the branchings exceeding 1~\%, 30~M are the decays
 K$^-\rightarrow \mu^-\pi^{\circ}\nu~($K$_{\mu3}$) and 20~M are the decays
 K$^-\rightarrow \pi^-\pi^{\circ}\pi^{\circ}~($K$_{\pi3}$).
 Signal efficiency 
  has been estimated using 5.7~M MC
  events of the radiative $K_{\mu3}$~ decay weighted with the matrix element, calculated
  in the leading approximation (up to terms of $O(p^4)$) of chiral perturbation 
  theory\cite{braga1,biji}
    
 Some information on the data
 processing and reconstruction procedures is given in
 \cite{ref2,ref3,ref4,ref2n,ref3n},
 here we briefly mention the details relevant for  present analysis.
 
 The muon identification (see \cite{ref3,ref3n}) is based on the information from the
  SP$_1$  and the HC. The energy deposition in the SP$_1$ is required to be compatible
  with the MIP signal in order to suppress charged pions and electrons. The sum of
  the signals in the HC cells associated with charged track is  required to
  be compatible with the MIP signal.  The muon selection is further enhanced by
  the requirement that the ratio $r_3$
   of the HC energy in last three
  layers to the total HC  energy  exceeds 5~\%. The used cut values are
  the same as  in~\cite{ref3n}.
  
   Events with one reconstructed charged track and three reconstructed
   showers in the calorimeter SP$_1$ are selected. We require the effective
   mass m$(\gamma\gamma)$ to be within $\pm 40$~MeV/c$^2$ from m$_{\pi^{\circ}}$. In the
   following analysis the central $\pm 20$~MeV/c$^2$ band is used for signal search
   and the side bands $95-115$~MeV/c$^2$ and $155-175$~MeV/c$^2$ are used  for background
   studies.      We require
   also the reconstructed $z$-coordinate of the vertex to be
   below 1650~cm.    183672 events have been selected and written to
   miniDST's using the above cuts with relaxed cut on $r_3$ to be above 1~\%. 
   
\section{ Evidence for signal and measurements of the branching ratios} 
  
  A set of cuts is developed  to suppress  
 backgrounds to the  K$_{\mu  3\gamma}$ decay and/or to do data cleaning:
 
 0) We select events with
  good charged track having two reconstucted ($x-z$ and $y-z$) projections and the number of
 hits in the MH below 5. We require also that the missing mass squared to the
  ($\mu^- \pi^{\circ} \gamma$)~system 
    abs(m$^2 ( \mu^- \pi^{\circ} \gamma)) < 0.05 $~(GeV/c$^2$)$^2$. 50804 events
    have survived these cuts.
 
 1) Events with the reconstructed vertex inside the interval
  $ 400 < z < 1600$~cm are selected.
  
 2) The measured missing energy $E_{mis}= E_{beam}-E_{\mu}-E_{\pi^{\circ}}-E_{\gamma}$ is
  required to be above zero.
  
 3)  We require the effective mass M$(\gamma\gamma)$ to be within $\pm 20$~MeV/c$^2$ from
  m$_{\pi^{\circ}}$.
  
 4) We require also that the missing mass squared to the $\pi^-\pi^{\circ}$~system is below
  0.025~(GeV/c$^2$)$^2$ ( m$^2(\pi^-\pi^{\circ})<0.025 $). 
  
 5)        The events with missing momentum pointing to the
    SP$_1$ working aperture are selected in order to suppress possible
 $\pi^-\pi^{\circ}\gamma$ background ( $6<r<60$~cm, here $r$ is the distance between
 the impact point of the missing momentum and the SP$_1$ center in the SP$_1$ transverse 
 plane).
  
 6) We require the photon energy E$^*_{\gamma}$ in the kaon rest frame to be below
   60~MeV.
  
 The remaining $K_{\pi2}$ decays are  suppressed by requirements:

 7) $ \cos (\theta) > -0.96$ , where $\theta$ is the angle between
 $\pi^-$ and $\pi^{\circ}$ in the kaon rest frame;

 8) $\varphi < 3.0$, where $\varphi$ is the angle between $\pi^-$
 and $\pi^{\circ}$ in the laboratory frame in the plane perpendicular
 to the beam momentum.
 
 9) We require also the absence of the signal above the threshold in the calorimeter SP$_2$.
 
  We look for a signal in the distributions over the effective mass
  M$(\mu^-\pi^{\circ}\gamma\nu)$, where $\nu$ four-momentum is calculated using the
   measured missing momentum and assuming m$_\nu=0$, and in the distributions of the
  missing mass squared to the $(\mu^- \pi^{\circ} \gamma)$-system,
   m$^2 ( \mu^- \pi^{\circ} \gamma )$.  
   Effective mass spectra for cut levels 1, 4, 6 and 9 are shown in Fig.~2. 
    These spectra show the evidence for peak   at m$_K$
   after the cut  on the photon  energy in the kaon rest frame.       
\begin{figure}
\epsfig{file=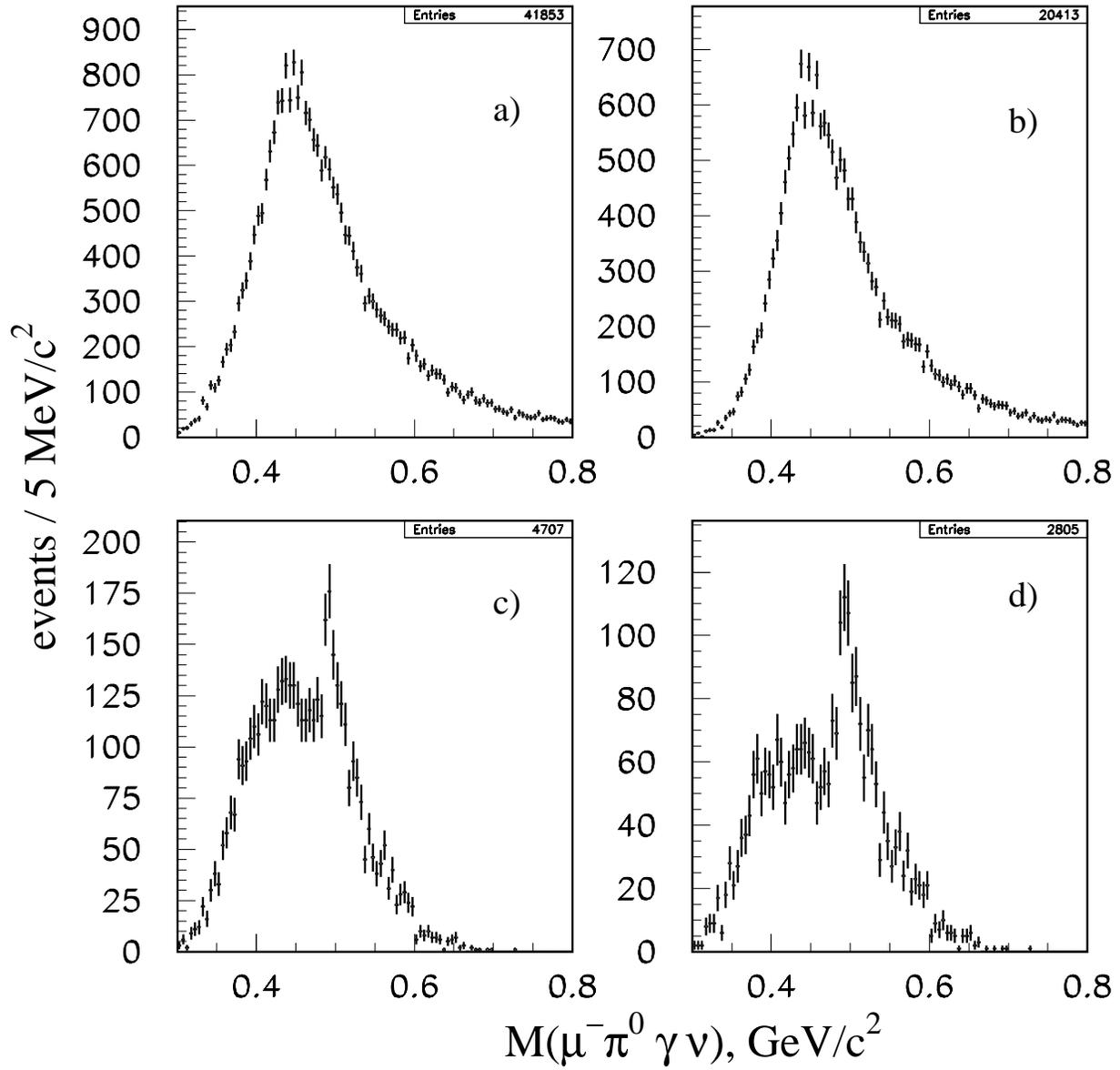,width=17.5cm}
\caption{ Effective mass M$(\mu^-\pi^{\circ}\gamma\nu$)
for the cut levels 1, 4 , 6 and 9 respectively.}
\end{figure}

\subsection{ The region below 30~MeV.}

 We have found that the signal is clearly seen for E$^*_{\gamma}<30$~MeV and 
 the background in this region is dominated by K$_{\mu3}$ decays (with an 
 accidental extra photon)
 and K$_{\pi3}$ decays. The main MC sample of 100~M events (with the natural mixture
 of the dominant decay modes) has been found to be insufficient for estimates of the
 background shapes, therefore specialized MC samples of 20~M K$_{\pi3}$ and 
 30~M K$_{\mu3}$ events have been used. The background has been divided into
 three contributions:
 
  1) Non-$\pi^{\circ}$ contribution has been estimated using tails of the
   M$(\gamma\gamma)$ distribution for real data, see Fig.~3 . 
\begin{figure}
\epsfig{file=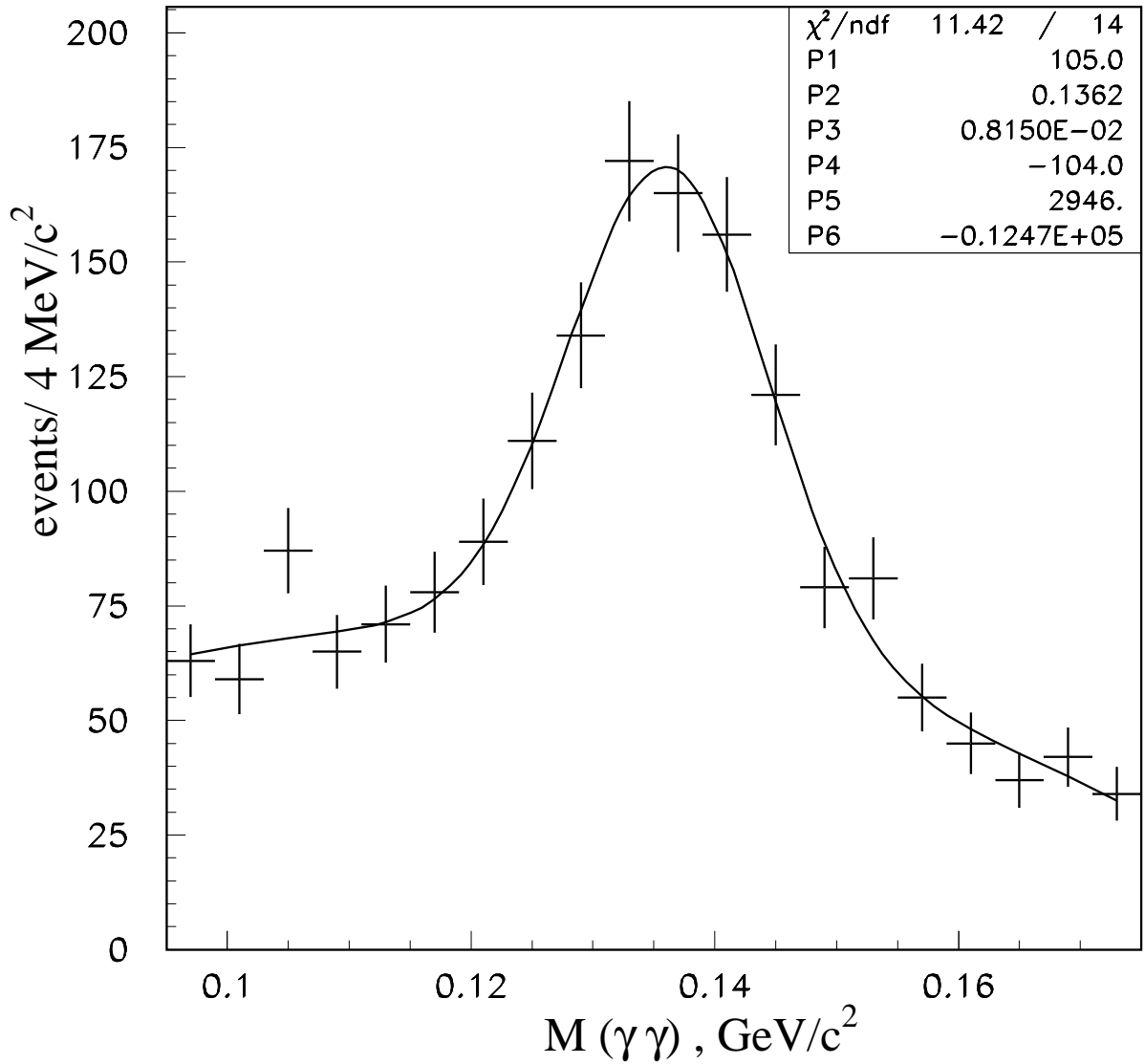,width=17.5cm}
\caption{ Effective mass M$(\gamma\gamma)$
 for events with E$^*_{\gamma}<30~$MeV at the cut level 9. Solid line
 shows parametrization by the sum of the Gaussian and polynomial background.}
\end{figure}

 2) K$_{\pi3}$ contribution has been approximated by the 
 form given by specialized MC sample, its
 normalization has been fixed using the observed K$_{\pi3}$ signal in the
  m$^2(\pi^-\pi^{\circ})$ distribution for selected events. 
  
3) K$_{\mu3}$ contribution has been approximated by the form from specialized
 MC sample, its normalization has been kept free.
 
 The shapes for all three background
 contributions have been found using the histogram smoothing
 by the HQUAD routine from the HBOOK package~\cite{hquad}. 
 The signal has been parametrized by the sum of two Gaussians with widths and relative
 fractions fixed at the values given by the signal MC sample.
\begin{figure}
\epsfig{file=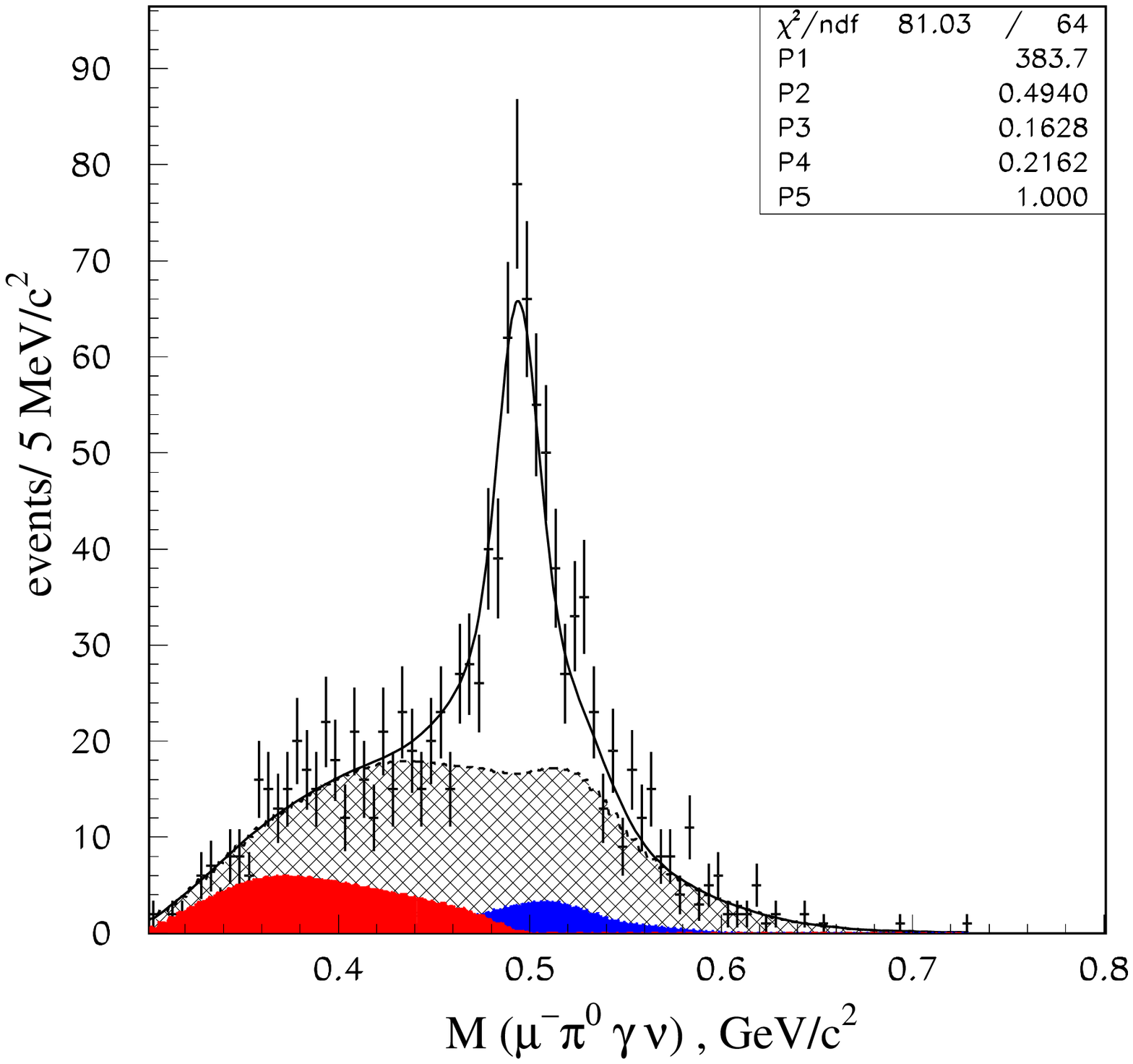,width=17.5cm}
\caption{ Effective mass M$(\mu^-\pi^{\circ}\gamma\nu)$ for events with
  $5 < $E$^*_{\gamma}<30~$MeV. Cross-hatched area shows the summary background,
 left(red) bump shows K$_{\pi3}$ contribution, right(blue) bump shows K$_{\mu3}$
 contribution. First parameter here and in the following figures is the number of
 signal events, second parameter is the position of the peak, third parameter is the
 normalization factor for K$_{\pi3}$ contribution, fourth parameter is the
 normalization factor for K$_{\mu3}$ contribution and fifth parameter is the
 normalization factor for non-$\pi^{\circ}$ contribution.} 
\end{figure}
\begin{figure}
\epsfig{file=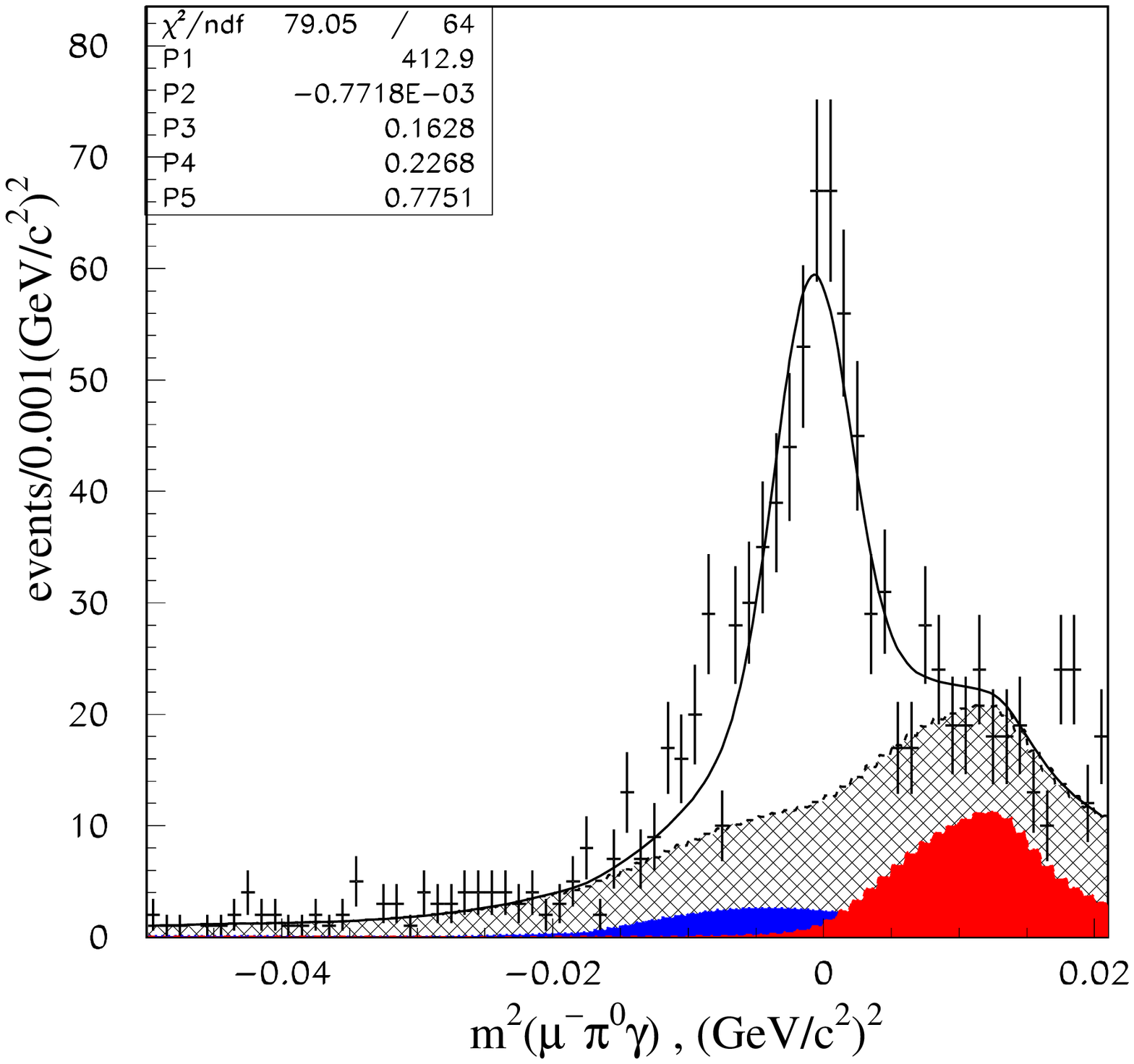,width=17.5cm}
\caption{ Missing mass squared m$^2(\mu^-\pi^{\circ}\gamma)$ for events with
 $5 < $E$^*_{\gamma}<30~$MeV. Cross-hatched area shows the summary background,
 right(red) bump shows K$_{\pi3}$ contribution, left(blue) bump shows K$_{\mu3}$
 contribution.}
\end{figure}

 Results of the fits are illustrated respectively in Fig.~4 and Fig.~5 for the
 distributions over M$(\mu\pi\gamma\nu)$ and m$^2 (\mu\pi\gamma)$. First parameter here (and 
 in the following) is
 the number of observed events, second parameter is the position of the peak , and
 three last parameters are the respective normalization factors
  of the K$_{\pi3}$, K$_{\mu3}$ and
 non-$\pi^{\circ}$ contributions.
 
 The number of observed events is equal to $ 383.7 \pm 40.9 $ in Fig.~4 and
  to $ 412.9 \pm 36.2 $ in Fig.~5. The difference between these two values (29.2) is our
  estimate of the systematics caused by the imprecise knowledge of the backgrounds.
  The results of fits with the polynomial parametrization of the background lie
  also within this uncertainty.
 
  The K$_{\mu3}$ decay has been used for the normalization.
 The number of K$_{\mu3}$ events in the region 400$ < z < 1600$~cm, corrected for
  the geometrical acceptance and the measurement efficiency,
  has been found using
 the cuts described in ~\cite{ref3n}. It is equal to 
  $ N($K$_{\mu3}) =5536000$. Independent normalization of  using K$_{\pi2}$
  decays gives the branching ratio lower by 6.2\%. The origin of this difference is explained 
  mainly by the trigger bias.
  This difference has been taken into
  account in our final estimates of the systematic uncertainties.
  
 The signal efficiency has been found from the signal MC  weighted with
 the matrix element, calculated within
 O$(p^4)$ ChPT approximation. It is equal to 2.6~\%. The signal efficiency
 has been calculated using the cut E$^*_{\gamma}> 5$~MeV, the cut
 value is our detection threshold explained by the beam momentum and the
 threshold in the energy of the SP$_1$ showers equal to $\approx 0.5$~GeV.
 
  The measured branching ratio is equal to  
  BR = ($8.82 \pm 0.94($stat$) \pm 0.86 ($syst$)) \times 10^{-5}$.
  This should be compared with theoretical prediction of 6.86$\times 10^{-5}$.  
 The ratio R=BR(K$_{\mu3\gamma})$/BR(K$_{\mu3})$ is equal to 
  R = ($2.70 \pm 0.29($stat$) \pm 0.26($syst$)) \times 10^{-3}$.
  This should be compared with theoretical prediction of 2.1$\times 10^{-3}$. In the transformations
  from the ratio R to the branching ratio we use the branching ratio
   BR(K$_{\mu3})=3.27~\%$~\cite{PDG}.
         
\subsection{Asymmetries for the region $5<$E$^*_{\gamma}<30$~MeV.}

  For this region we have measured the asymmetry of photon emission towards
   muon direction in  the kaon rest frame
   $A(\cos \theta^*_{\mu\gamma})$ 
\begin{center}  
   A$(\cos \theta^*_{\mu\gamma} ) = \frac{N(\cos\theta^*_{\mu\gamma} >0) - 
   N(\cos\theta^*_{\mu\gamma} <0)}
   {N(\cos\theta^*_{\mu\gamma} >0) + N(\cos\theta^*_{\mu\gamma} <0)}$
\end{center}   
 and the asymmetry in $\xi$, A$(\xi)$. The effective mass spectra are shown 
 in Fig.~6 for positive and negative $\cos\theta^*_{\mu\gamma}$ separately
 and in Fig.~7 for positive and negative $\xi$.
  
  The asymmetry in $\cos \theta^*_{\mu\gamma}$ is equal to $0.09 \pm 0.14$, 
  this value is below the
  theoretical expectation, equal to 0.35, by 2 standard deviations.
  
  The asymmetry in $\xi$ is equal to $-0.03 \pm 0.13$. Of course our statistics is insufficient to
  test the theoretical predictions~\cite{braga1,braga2}.  
\begin{figure}
\epsfig{file=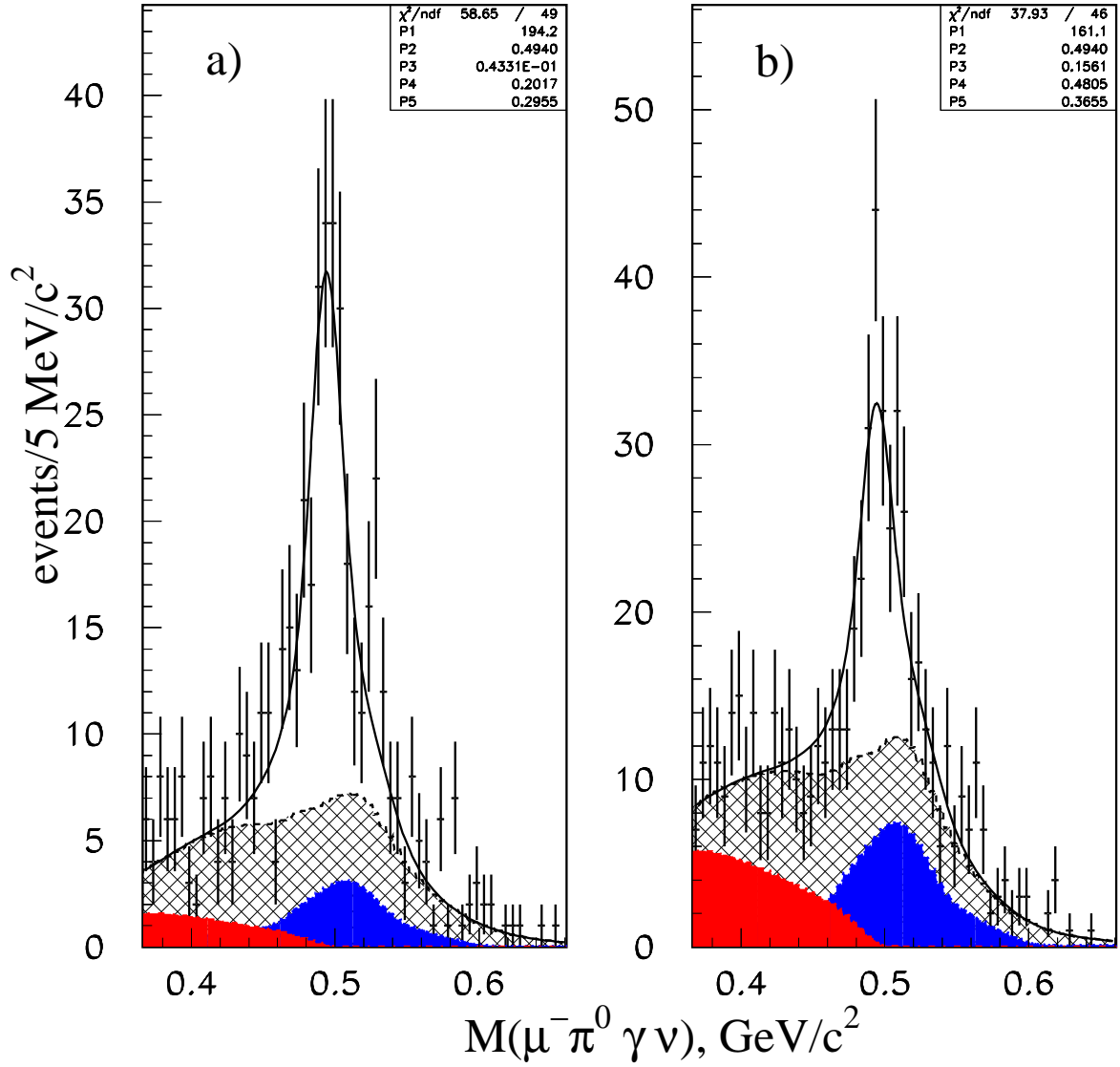,width=17.5cm}
\caption{ Effective mass M$(\mu^-\pi^{\circ}\gamma\nu)$ spectra for: 
a) $\cos (\theta^*(\mu\gamma))>0$ and b) $\cos (\theta^*(\mu\gamma))<0$. The background
 notations are the same as in Fig.4.}
\end{figure}
\begin{figure}
\epsfig{file=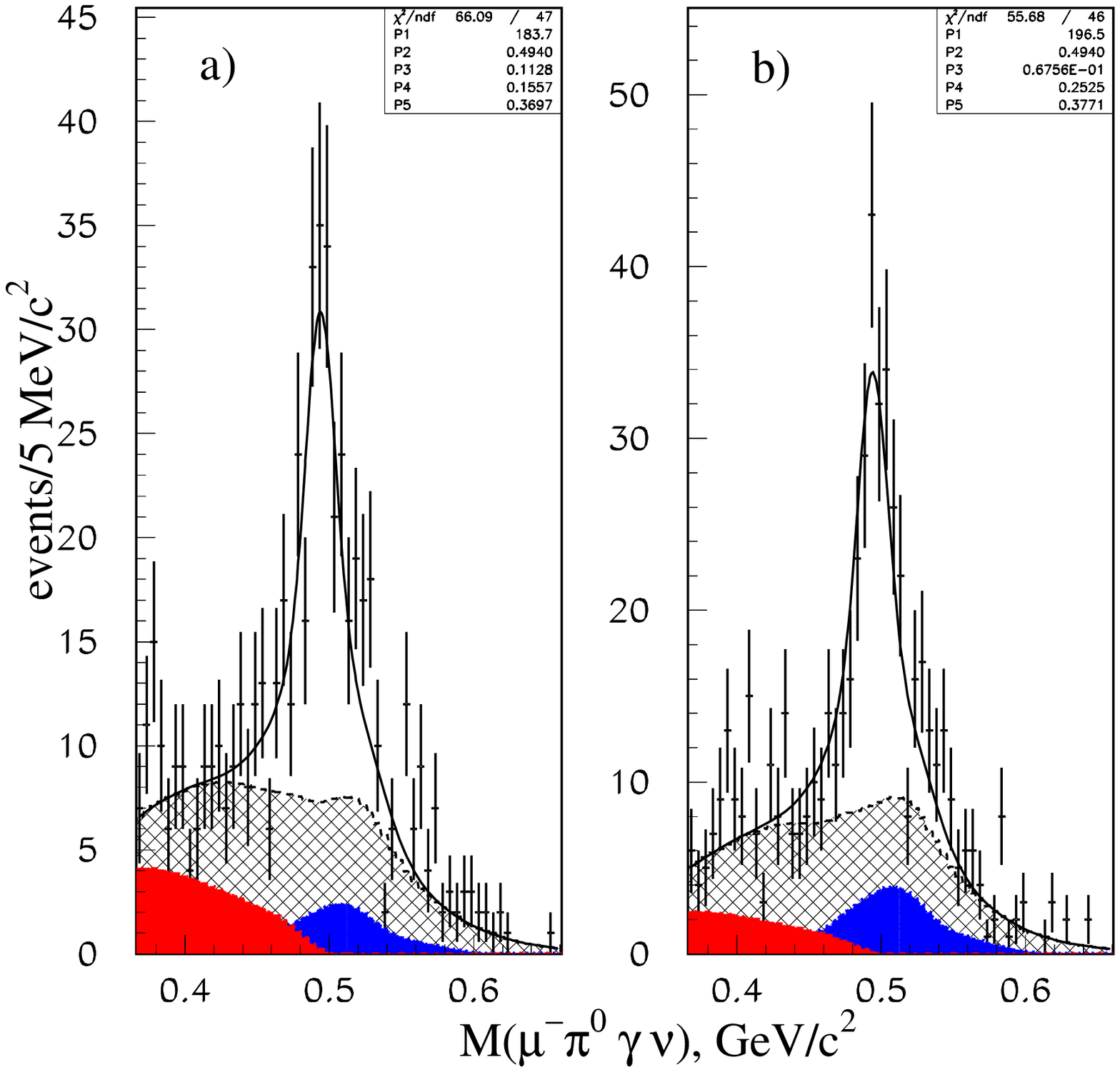,width=17.5cm}
\caption{ Effective mass M$(\mu^-\pi^{\circ}\gamma\nu)$ spectra for: 
a) $\xi>0$ and b) $\xi<0$. The background notations are the same as in Fig.4.}
\end{figure}

\subsection{ The region $30<$E$^*_{\gamma}<60$~MeV.}

  For this region we see strong K$_{\pi3}$ background and residual K$_{\pi2}$ background.
  These backgrounds have been   suppressed  by additional cut 0.1$<p^* (\pi^-)<0.185$~MeV/c.
  The parametrization is illustrated in
  Fig.~8  for the effective mass M$(\mu^-\pi^{\circ}\gamma\nu)$ spectrum.
  The signal efficiency was found to be 6.30~\%.  
  The number of observed events is equal to  
  $152.7 \pm 23.06($stat$) \pm 32.2($syst$)$. The systematics in the number of observed events
  has been calculated in the same way as in the Section 3.1.
  The branching ratio  
  BR = ($1.46 \pm 0.22($stat$) \pm 0.32($syst$)) \times 10^{-5}$, is compatible
  with the theoretical expectation $1.53 \times 10^{-5}$.  
  The ratio 
  R = ($4.48 \pm 0.68($stat$) \pm 0.99($syst$)) \times 10^{-4}$, is 
  compatible
  with the expectation $4.67 \times 10^{-4}$.    
\begin{figure}
\epsfig{file=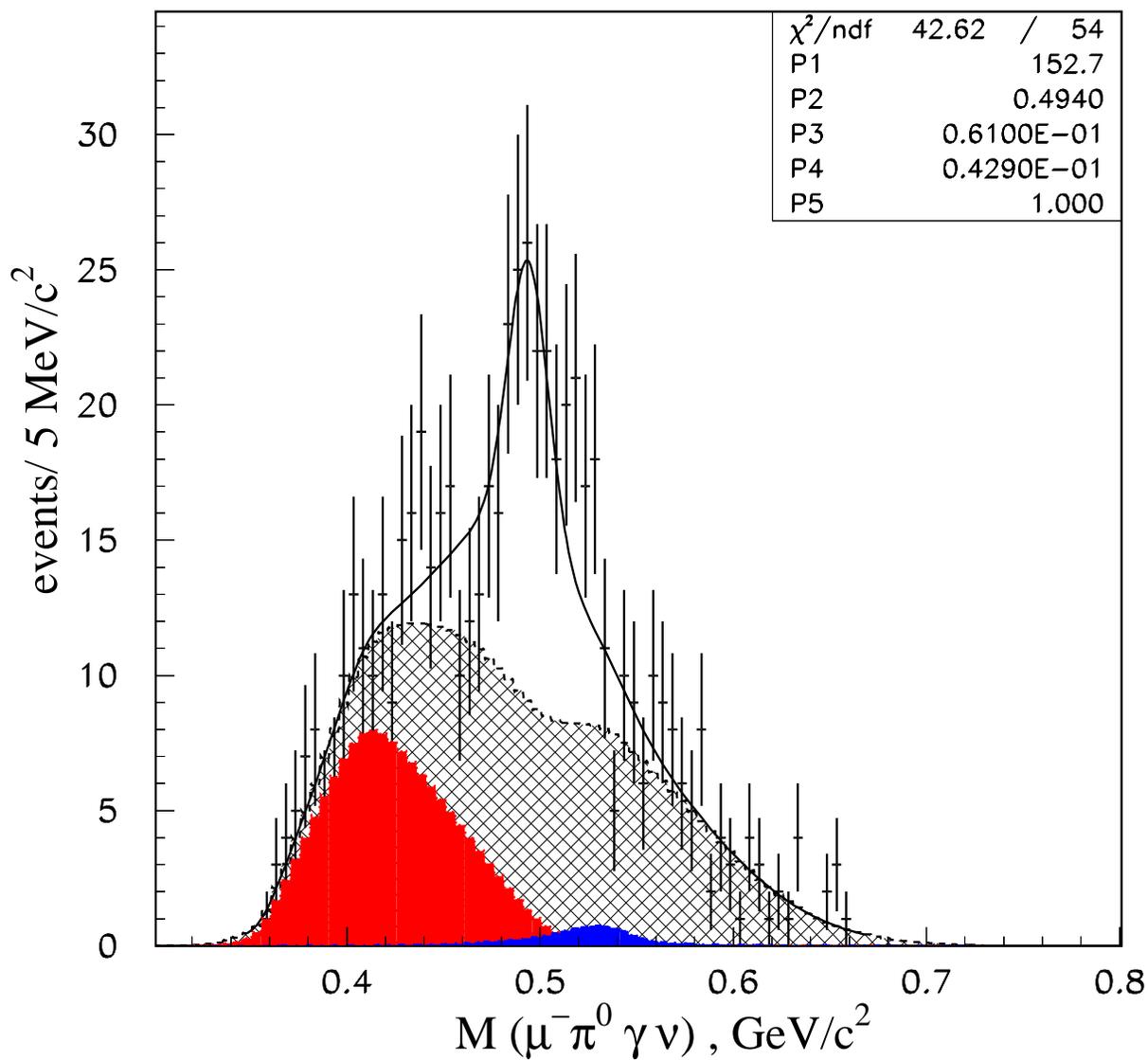,width=17.5cm}
\caption{ Effective mass M$(\mu^-\pi^{\circ}\gamma\nu)$ spectrum 
 for $30<$E$^*_{\gamma}<60$~MeV.
 Cross-hatched area shows the  background. Solid areas, left(red) and
 rigth(blue), show K$_{\pi3}$ and K$_{\mu3}$ contributions respectively.}
\end{figure}

\newpage   	        
\section{ Conclusions}

  Our conclusions are the following.
  
$\bullet$ First observation of the radiative kaon decay K$^-_{\mu3\gamma}$
 is presented.
 
$\bullet$ The measured ratio R = BR(K$_{\mu3\gamma})/$BR(K$_{\mu3})$ for the region
 $5< $E$^*_{\gamma}<30$~MeV is equal to
  0.270$\pm 0.029($stat$) \pm 0.026($syst$)\%$. This is consistent with theoretical 
  prediction equal to 0.21~\%.

$\bullet$ The measured ratio R for the region $30< $E$^*_{\gamma}<60$~MeV is equal to
 ($4.48 \pm 0.68($stat$) \pm 0.99($syst$)) \times 10^{-4}$, this value is compatible
 with theoretical prediction equal to $4.67 \times 10^{-4}$.

$\bullet$ The measured asymmetry in the T-odd variable $\xi$ for the region
 $5< $E$^*_{\gamma}<30$~MeV is  equal to $-0.03 \pm 0.13$.
 
$\bullet$ The measured asymmetry in the $\cos \theta^*_{\mu\gamma}$ is equal to
 $0.093 \pm 0.141$, this value is two standard deviations away from the theoretical
 prediction equal to 0.354. 
   
 The work 
 is  supported in part by the RFBR grant N03-02-16330(IHEP group) and
 RFBR grant N03-0216135(INR group). We are indebted to V.~Braguta for
 giving us routine for O$(p^4)$ ChPT matrix element calculations.


\end{document}